\documentclass[11pt,a4paper,nofootinbib,showpacs,preprint,utf8]{article}
\usepackage{jheppub}
\pdfoutput=1
\usepackage{amsmath,amssymb}
\usepackage{amsfonts}
\usepackage{epsfig}
\usepackage{subcaption}

\usepackage{graphicx}
\usepackage{bbm}
\usepackage{cancel}
\usepackage{pifont}
\usepackage[usenames,dvipsnames]{xcolor}
\definecolor{warmblack}{rgb}{0.0, 0.26, 0.26}
\definecolor{mediumtealred}{rgb}{0.0, 0.33, 0.71}
\usepackage{slashed}
\usepackage[compat=1.1.0]{tikz-feynman}
\usepackage{tikzsymbols}
\usepackage{comment}
\usepackage{appendix}
\usepackage[normalem]{ulem}
\usepackage{float}

\title{Dark-Portal Leptogenesis in a Non-Holomorphic Modular Scoto-Seesaw Model}
\author[a]{Salah Nasri}
\author[b]{, Labh Singh}
\author[b]{, Tapender}
\author[b]{, Surender Verma}

\affiliation[a]{Department of physics, United Arab Emirates University, Al-Ain, UAE} 

\affiliation[b]{Department of Physics and Astronomical Science, Central University of Himachal Pradesh, Dharamshala, Himachal Pradesh 176215, India}

\emailAdd{snasri@uaeu.ac.ae}
\emailAdd{sainilabh5@gmail.com}
\emailAdd{tapenderphy@gmail.com}
\emailAdd{s\_7verma@hpcu.ac.in}

\abstract{This work explores the neutrino phenomenology of the scotoseesaw model under non-holomorphic $A_4$ modular flavor symmetry providing a non-SUSY framework for realization of the modular symmetry. To prevent mixing between the beyond standard model fields associated with the tree and loop-level neutrino mass contributions, we assign even and odd modular weights to these sectors, respectively.  The physical allowed ranges of oscillation parameters are used to identify the viable region of modulus parameter $\tau$ in its fundamental domain. With the complex modulus $\tau$ serving as the unique source of CP violation (all other parameters are real) the framework realizes successful low-scale leptogenesis through CP-violating decays of the lightest right-handed neutrino into Standard Model leptons and the Higgs boson. The requisite CP asymmetry arises from one-loop diagrams involving dark-sector states, obviating the need for degenerate mass spectra and thereby circumventing the usual resonant leptogenesis mechanism. The observation of a long-lived charged particle ($\eta^{\pm}$) in collider experiments would offer compelling evidence for the inert scalar sector of the model and provide a crucial experimental hint on the dark-sector assisted generation of neutrino masses and leptogenesis.}

\keywords{Modular symmetry; Radiative neutrino mass; Phenomenology; Leptogenesis; Dark matter.}
\makeatletter
\gdef\@fpheader{}
\makeatother

\begin{document}
\maketitle

\section{Introduction}\label{section1} 
The confirmed observation of neutrino oscillations provides definitive evidence that neutrinos are massive and undergo flavor mixing phenomena that necessitate physics beyond the Standard Model (SM) \cite{Super-Kamiokande:1998kpq, Ahmad:2002jz,An:2012eh,Abe:2011fz,KamLAND:2002uet,SNO:2001kpb}. A fundamental theoretical challenge is the absence of right-handed neutrino counterparts in the SM, which precludes the generation of neutrino mass \textit{via} the conventional Higgs mechanism. While the dimension-five Weinberg operator \cite{Weinberg:1979sa} offers a viable framework for neutrino mass generation, its ultraviolet completion and the specific flavor structure of neutrino sector remain open questions. Consequently, exploring beyond the standard model (BSM) scenarios is crucial for understanding the origin of neutrino mass. A wide spectrum of models has been proposed to align with oscillation data, including various implementations of the seesaw mechanism, radiative mass generation, and models involving extra dimensions \cite{Cai:2017jrq,Gu:2020nic,Bonnet:2012kz,Liao:2010ku,Chala:2021juk,Li:2023ohq,Du:2022vso,Ma:2006km,Tapender:2025xoj,Tapender:2024ktc,Tapender:2023kdk,Singh:2022nmk,Kashav:2021zir,Priya:2025wdm,Singh:2024imk,Singh:2023eye,Singh:2025jtn,Kashav:2023tmz,Verma:2018lro,Ng:2003rv,Arkani-Hamed:1998wuz,Neubert:2000kj,Asaka:2019vev,Ding:2022bzs,Behera:2022wco,Pathak:2025zdp,Kashav:2022kpk,Mishra:2022egy,Marciano:2024nwm,Behera:2020sfe,Mohanta:2023tzf,Abhishek:2025ety}
. A common feature in many of these frameworks is the introduction of sterile neutrinos: SM gauge singlets that act as right-handed partners. These sterile neutrinos couple to the active left-handed neutrinos \textit{via} Yukawa interactions. The masses and coupling strengths of these sterile states can span many orders of magnitude, leading to diverse phenomenological implications. For instance, within the canonical type-I seesaw, the observed sub-eV neutrino masses naturally point to a very high mass scale for the right-handed neutrinos, around $10^{15}$ GeV, which is inaccessible to current experiments. However, alternative realizations of the seesaw mechanism—such as the inverse, linear, or extended seesaw, often constructed with discrete or continuous symmetries \cite{Nogueira:2025hch,Kang:2006sn,CarcamoHernandez:2024edi,Dias:2011sq,Ma:2009kh,Batra:2023mds,Das:2017ski} can feature sterile neutrino masses at the TeV scale, rendering them potentially testable at current or future colliders. To date, neutrino oscillation experiments have precisely measured two mass-squared differences associated with atmospheric and solar neutrino oscillations. The existence of these two distinct mass scales may itself be a clue, potentially indicating that the masses for the second and third  generations originate from different underlying mechanisms. 
\\

\noindent To validate our assertion, we employ a recently proposed idea of the non-holomorphic modular symmetry, a non-SUSY realization of the holomorphic modular symmetry \cite{Qu:2024rns}. This approach is rooted in the construction of non-holomorphic modular forms, specifically polyharmonic Maaß forms. Unlike their holomorphic counterparts these forms are invariant under the modular group but are not required to be holomorphic, instead, they satisfy eigenfunction conditions under a Laplace-type differential operator. This expanded functional space incorporates both holomorphic and non-holomorphic components and accommodates modular forms of both positive and negative weights. Consequently, it enables the construction of more flexible and phenomenologically viable Yukawa coupling structures. Crucially, the rigid constraint of modular invariance is preserved, ensuring the model's predictive power while circumventing the limitations intrinsic to a purely holomorphic framework. Several works pertaining to the neutrino mass model building have already been done in the non-holomorphic modular symmetry \cite{Nomura:2024atp, Nomura:2024vzw, Nomura:2024nwh, 
Nomura:2025ovm, Nomura:2025raf, Nomura:2025bph, 
Kang:2024jnp, Ding:2024inn, Li:2024svh, Okada:2025jjo, Kobayashi:2025hnc, Loualidi:2025tgw, Zhang:2025dsa, Priya:2025wdm, Nomura:2024ctl, Abbas:2025nlv, Li:2025kcr, Dey:2025zld,Nanda:2025lem,Kumar:2025bfe}.
\\

\noindent Further, the observed cosmic matter-antimatter asymmetry remains unexplained in the SM. Leptogenesis is a leading theoretical solution, proposing that this imbalance originated in the lepton sector \cite{Fukugita:1986hr,Davidson:2008bu,Buchmuller:2005eh,Fong:2012buy,Pilaftsis:2009pk}. It typically involves the out-of-equilibrium, CP-violating decay of heavy right-handed neutrinos in the early universe. This lepton asymmetry is subsequently converted into the observed baryon asymmetry \textit{via} sphaleron processes. A notable feature of leptogenesis is its connection to the low-energy neutrino phenomenology through heavy right-handed degrees of freedom in the seesaw framework. The modular symmetry framework offers a highly predictive setting for leptogenesis, as the Yukawa couplings become functions of the complex modulus $\tau$, which in turn acts as a natural source of CP asymmetry. Leptogenesis within modular-invariant theories has been extensively studied in holomorphic formulations~\cite{Asaka:2019vev,Ding:2022bzs,Behera:2022wco,
Pathak:2025zdp,Kashav:2022kpk,Mishra:2022egy,Marciano:2024nwm,Behera:2020sfe,Mohanta:2023tzf,Abhishek:2025ety,Kashav:2021zir,Singh:2024imk} and some  non-holomorphic extensions~\cite{Priya:2025wdm,Nanda:2025lem,Kumar:2025bfe}.
\\

\noindent Motivated by this, we explored the \textit{scoto-seesaw} framework \cite{Kubo:2006rm,Rojas:2018wym} where one mass square difference originates at the seesaw level and another at the one-loop (scotogenic) level. To forbid mixing between the tree-level and loop-level BSM sectors, distinct modular weights ($\kappa$) are assigned: tree-level fields carry even $\kappa$, whereas loop-level fields are odd under $\kappa$. In this setup, the heavy right-handed neutrino $N_1$, responsible for the tree-level seesaw contribution, decays into the SM leptons and Higgs doublets. The dark sector, consisting of a chiral singlet fermion $f$, an inert scalar doublet, and a real scalar singlet $s-$ all odd under $\kappa-$ plays a multifaceted role: (i) it induces the one-loop radiative neutrino mass contribution, (ii) it mediates loop-level corrections to the $N_1$ decay amplitude whose interference with the tree-level process generates a non-zero $CP$ asymmetry, and (iii) it offers a viable dark matter candidate within the model. The interaction $\mu_{1}\,s\,\eta^{\dagger}H$, together with the Yukawa structure involving the fermion multiplet $f$ and the singlet $s$, directly enters the one-loop diagrams responsible for leptogenesis. Consequently, the presence of $s$ and the other BSM fields is not incidental but essential, as they generate the loop-induced CP asymmetry, modify the washout structure, and allow successful leptogenesis at a comparatively lower scale without requiring near-degenerate heavy neutrinos.
\\

\noindent The paper is organized as follows. In Section~\ref{section2}, we present the details of the model framework. Section~\ref{section3} is devoted to explain the procedure involved in numerical estimation of neutrino masses and mixing parameters. In Section~\ref{section4} we discuss the charged lepton flavor violation in the model. A discussion of  dark-sector assisted leptogenesis, together with brief comments  on the dark matter implications, is presented Section~\ref{section5}. Finally, conclusions are summarized in Section~\ref{section6}.

\section{Model and Formalism}
The modular symmetry-based Scoto-Seesaw model is given in the Table \ref{modelpart}. The basic algebra of the modular symmetry has been given in Appendix \ref{sec:non-holomorphic-modular}. To ensure that there is no mixing between the tree and loop level BSM fields, we have assigned even and odd weights to the tree and loop level fields.
\label{section2}

\begin{table}[t]
    \centering
\begin{tabular}{|c|c|c|c|}
\hline
Fields & $SU(2)_L$ & $A_4$ & $\kappa$ \\
\hline
\hline
$L$ & 2 & 3 & 2 \\

$e_{R},\mu_{R},\tau_{R}$ & 1 & $1,\;1^\prime,\;1^{\prime \prime}$ & -2 \\

$H$ & 2 & 1 & 0 \\

$N_1$ & 1 & 1 & 2 \\

$N_2$ & 1 & $1^{\prime}$ & 2 \\

$f$ & 1 & 1 & 1 \\

$\eta$ & 2 & 1& 1\\

$s$ & 0 & 1& -1\\
\hline
\end{tabular}
\caption{The particle content and charge assignments of the Scoto-Seesaw model under $SU(2)_L$ and $A_4$ non-holomorphic modular symmetry.}
\label{modelpart}
\end{table}
\noindent The relevant scalar potential of the model is given by
\begin{equation}
\begin{aligned}
V_{\text{tree}} = &\; \mu_{H}^2 |H|^2 + \mu_{\eta}^2 |\eta|^2 
+ \lambda_1 |H|^4 + \lambda_2 |\eta|^4 + \lambda_3 |H|^2 |\eta|^2 
+ \lambda_4 |\eta^\dagger H|^2 + \tilde{\lambda}_5 [(\eta^\dagger H)^2 + \text{h.c.}] \\
& + \frac{\mu_{s}^2}{2} s^2 + \lambda_7 s^4 + \lambda_8 s^2 |H|^2 
+ \lambda_9 s^2 |\eta|^2 + (\mu_1 s H^\dagger \eta + \mu_1^* s \eta^\dagger H).
\end{aligned}
\end{equation}
Here $H$ is the SM Higgs doublet\footnote{ Although $\mu_1$, in general, can be complex, we take it as real in our study. It is to be noted that most of the terms in the scalar potential remain invariant under modular symmetry as given in \cite{Nomura:2024vzw}.} and $\tilde{\lambda}_5=\lambda Y_1^{(2)}$. Also, the scalar dark sector comprises of Higgs-type scalar doublet
\begin{eqnarray}
\eta= \begin{pmatrix}
        \eta^+ \\
        \frac{(\eta_R +i \eta_I)}{\sqrt{2}}
    \end{pmatrix},
\end{eqnarray}
and a real singlet scalar $s$. After spontaneous symmetry breaking, the charged and odd component of the $\eta$ acquire mass as
\begin{eqnarray}
m_{\eta^{+}}^2&=&\mu^{2}_{\eta}+\frac{1}{2}\lambda_3 v^2,\\
m_{\eta_I}^2&=& \mu^{2}_{\eta}+\frac{1}{2}(\lambda_3+\lambda_4-\tilde{\lambda}_5)v^2,
\end{eqnarray}
where $v$ is the vacuum expectation value (\textit{vev}) of the SM Higgs. In the $(\eta_R,s)$ basis, the squared mass matrix is non-diagonal and is given by
\begin{eqnarray}
M^2 =
\begin{pmatrix}
\dfrac{\partial^2 V}{\partial \eta_R^2} &
\dfrac{\partial^2 V}{\partial \eta_R \partial s} \\[8pt]
\dfrac{\partial^2 V}{\partial s \partial \eta_R} &
\dfrac{\partial^2 V}{\partial s^2}
\end{pmatrix}
=
\begin{pmatrix}
\mu^{2}_{\eta}+\frac{1}{2}(\lambda_3+\lambda_4-\tilde{\lambda}_5)v^2 & \mu_1 v \\
\mu_1 v & \mu^2_{s}+\frac{1}{2}\lambda_8 v^2
\end{pmatrix},
\end{eqnarray}
which can be diagonalized as
\begin{eqnarray}
\begin{pmatrix}
\eta_1 \\
\eta_2
\end{pmatrix}
=
\begin{pmatrix}
\cos\theta & \sin\theta \\
-\sin\theta & \cos\theta
\end{pmatrix}
\begin{pmatrix}
\eta_R \\
s
\end{pmatrix},
\end{eqnarray}
such that
\begin{eqnarray}
\tan{2\theta}=\frac{2 \mu_1 v}{\mu^{2}_{\eta}-\mu^2_{s}+\frac{1}{2}\left(\lambda_3+\lambda_4-\tilde{\lambda}_5-\lambda_8 \right)v^2}.
\end{eqnarray}
The mass of CP even scalars is given by
\begin{eqnarray}\nonumber
m_{\eta_1}^2
&=&\frac12\!\left[
\mu_{\eta}^2+\mu_{s}^2
+\frac12(\lambda_3+\lambda_4-\tilde{\lambda}_5+\lambda_8)v^2
+\sqrt{\left(\mu_{\eta}^2-\mu_{s}^2
+\frac12(\lambda_3+\lambda_4-\tilde{\lambda}_5-\lambda_8)v^2\right)^2
+4\mu_1^2 v^2}
\right],\\
\nonumber
m_{\eta_2}^2
&=&\frac12\!\left[
\mu_{\eta}^2+\mu_{s}^2
+\frac12(\lambda_3+\lambda_4-\tilde{\lambda}_5+\lambda_8)v^2
-\sqrt{\left(\mu_{\eta}^2-\mu_{s}^2
+\frac12(\lambda_3+\lambda_4-\tilde{\lambda}_5-\lambda_8)v^2\right)^2
+4\mu_1^2 v^2}
\right].
\end{eqnarray}
Using a binomial expansion\footnote{We considered $\lambda_8$ small to eradicate unwanted mixing between SM Higgs and singlet scalar $s$.}, the mass-squared eigenvalues of the inert scalars can be approximated as
\begin{equation}
m_{\eta_{1,2}}^2 \simeq \frac{1}{4} \Bigg[ 2 (\mu_\eta^2 + \mu_s^2) + v^2 (\lambda_3 + \lambda_4 - \tilde{\lambda}_5) 
\pm \sqrt{\big(v^2 (\lambda_3 + \lambda_4 - \tilde{\lambda}_5) - 2 \mu_s^2 + 2 \mu_\eta^2 \big)^2 + 16 \mu_1^2 v^2} \, \Bigg]
\label{darkmasses}
\end{equation}
The dark matter candidate in the model can be either a CP even ($\eta_1$,$\eta_2$) or a  CP odd ($\eta_I$) scalar, depending on the mass spectrum considered. 
\noindent The relevant Yukawa Lagrangian for the charged leptons,  based on the field content given in the Table~\ref{modelpart}  is given by
\begin{eqnarray}
\mathcal{L}\supset \alpha_1 Y_3^{(0)} \left(\bar{L}H e_{R}\right) + \alpha_2  Y_3^{(0)} \left(\bar{L}H \mu_{R}\right) + \alpha_3  Y_3^{(0)} \left(\bar{L}H \tau_{R}\right)
\end{eqnarray}
where $Y_3^{(0)}=(Y^0_1,Y^0_2,Y^0_3)$ is an $A_4$ triplet modular form with weight 0. From this Lagrangian,  charged lepton mass matrix can be written as


\[
M_\ell = v 
\begin{pmatrix}
\alpha_1 Y^0_1& \alpha_2 Y^0_3 &\alpha_3 Y^0_2 \\
\alpha_1 Y^0_3& \alpha_2 Y^0_2 &\alpha_3 Y^0_1 \\
\alpha_1 Y^0_2& \alpha_2 Y^0_1 &\alpha_3 Y^0_3
\end{pmatrix}= v\,\alpha_1
\begin{pmatrix}
Y^0_1 & r_{21}\,Y^0_3 & r_{31}\,Y^0_2 \\[4pt]
Y^0_3 & r_{21}\,Y^0_2 & r_{31}\,Y^0_1 \\[4pt]
Y^0_2 & r_{21}\,Y^0_1 & r_{31}\,Y^0_3
\end{pmatrix},
\]
where $r_{21}=\alpha_2/{\alpha_1}$ and $r_{31}=\alpha_3/{\alpha_1}$. The invariant Yukawa Lagrangian for neutrinos,  including the Dirac Yukawa interactions and Majorana mass terms for the right-handed neutrinos, is given by
\begin{eqnarray}
\mathcal{L} \supset \gamma \left(Y_{3}^{(4)} \bar{L} \tilde{H} N_1 + Y_{3}^{(4)} \bar{L} \tilde{H} N_2\right) + \kappa_{1}Y_{1}^{(4)}N_1 N_1 + \kappa_{2}Y_{1'}^{(4)}N_2 N_2,
\end{eqnarray}
the right-handed neutrino mass matrix is then given by
\begin{eqnarray}
M_R = \begin{pmatrix}
\kappa_1 Y_1^{(4)} & 0 \\
0 & \kappa_2 Y_{1'} ^{(4)}
\end{pmatrix},
\end{eqnarray}
and Dirac neutrino mass matrix is
\begin{eqnarray}
M_D = \gamma v\begin{pmatrix}
Y_1^{4} & Y_3^{4} \\
Y_3^{4} & Y_2^{4} \\
Y_2^{4} & Y_1^{4}
\end{pmatrix},
\end{eqnarray}
where $Y_3^{(4)}=(Y^4_1,Y^4_2,Y^4_3)$ is $A_4$ triplet modular form with weight 4.
The neutrino masses generated by Type-I seesaw is given by
\begin{eqnarray}\label{mnu1}
(m^{\nu})_{\text{tree}}=-M_{D}M_{R}^{-1}M_{D}^{T}.
\end{eqnarray}
Additionally, the model exhibits a scotogenic nature. The relevant Yukawa Lagrangian, which is important for the one-loop generation of neutrino mass can be expressed as
\begin{eqnarray}
L=\beta_L Y_3^{(4)} \bar{L} \tilde{\eta} f + \kappa_S Y_1^{(2)} f f + aY_1^{(2)} \bar{N}^c f s.
\end{eqnarray}
The loop-induced  neutrino mass matrix takes the form 
\begin{equation}
(m^{\nu}_{ij})_\text{loop} = F(m_{\eta_1}, m_{\eta_2}, m_{\eta_I}) M_f h_i h_j,
\end{equation}
where $M_f = \kappa_S Y_1^{(2)}$,  and  $F(m_{\eta_1}, m_{\eta_2},  m_{\eta_I})$ denotes the loop function
\begin{eqnarray}
F(m_{\eta_1}, m_{\eta_2}, m_{\eta_I})=\frac{1}{32\pi^2}\left[\cos{\theta}F(m_{\eta_1})+\sin{\theta}F(m_{\eta_2})-F(m_{\eta_I})\right],
\end{eqnarray}
with 
\begin{eqnarray}
F(m_x)=\frac{m_{x}^2}{M_{f}^2-m_{\eta_x}}\text{ln}\frac{m_x^2}{M_f^2}.
\end{eqnarray}
The neutrino mass matrix generated at the loop level is
\begin{equation}\label{mnu2}
(m^\nu)_{\text{loop}} = \beta^2_L M_f
\begin{pmatrix}
(Y^4_1)^2 & Y^4_1 Y^4_2 & Y^4_1 Y^4_3 \\
Y^4_1 Y^4_2 & (Y^4_2)^2 & Y^4_2 Y^4_3 \\
Y^4_1 Y^4_3 & Y^4_2 Y^4_3 & (Y^4_3)^2
\end{pmatrix}
F(m_{\eta_1}, m_{\eta_2}, m_{\eta_I}).
\end{equation}
Considering Eqns. (\ref{mnu1}) and (\ref{mnu2}), the total neutrino mass matrix is given by
\begin{eqnarray}
M^{\nu}=(m^{\nu})_\text{tree}+(m^\nu)_{\text{loop}}.
\label{neutrinomassmatrix}
\end{eqnarray} 
The next step is to ascertain the viability of the model to reproduce low energy neutrino phenomenology which will help to identify the allowed parameter space of, initially, free parameters. In order to get the predictions for the neutrino oscillation parameters, we will numerically diagonalize $M^{\nu}$ in the next section.
\section{Neutrino Masses and Mixing}\label{section3}

\begin{figure}
    \centering
    \begin{subfigure}[b]{0.4\textwidth} 
        \centering
        \includegraphics[width=\linewidth]{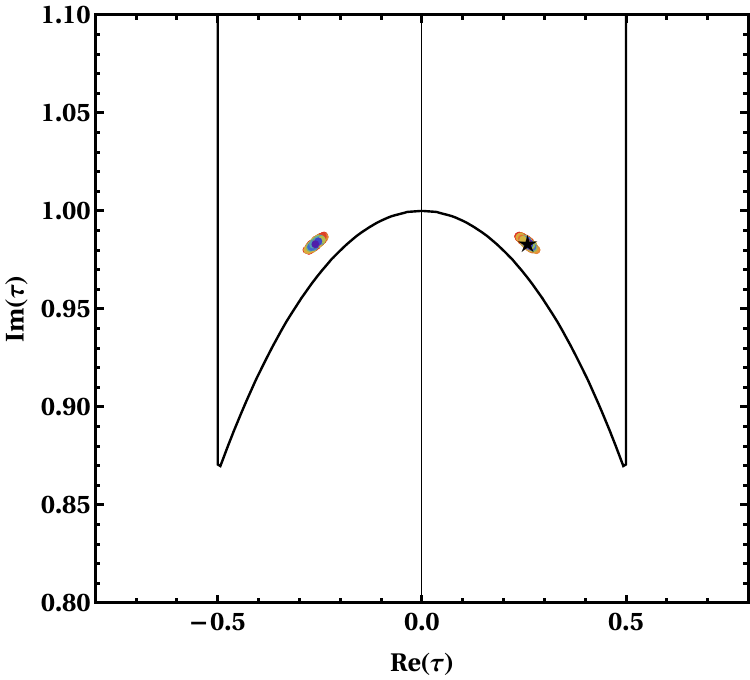}
        \caption{}
        \label{fig1}
    \end{subfigure}
    \hfill
    \begin{subfigure}[b]{0.4\textwidth}
        \centering
        \includegraphics[width=\linewidth]{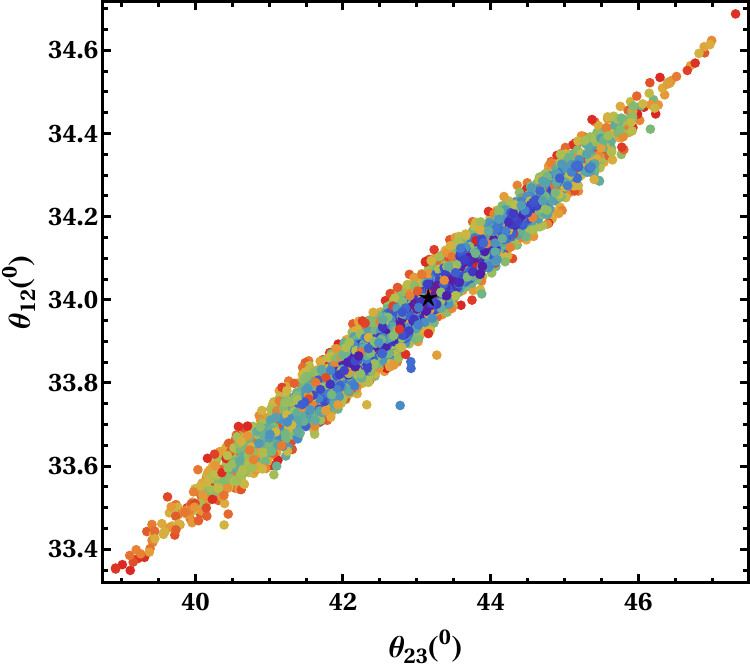}
        \caption{}
        \label{fig2}
    \end{subfigure}
    \hfill
    \begin{subfigure}[b]{0.4\textwidth}
        \centering
        \includegraphics[width=\linewidth]{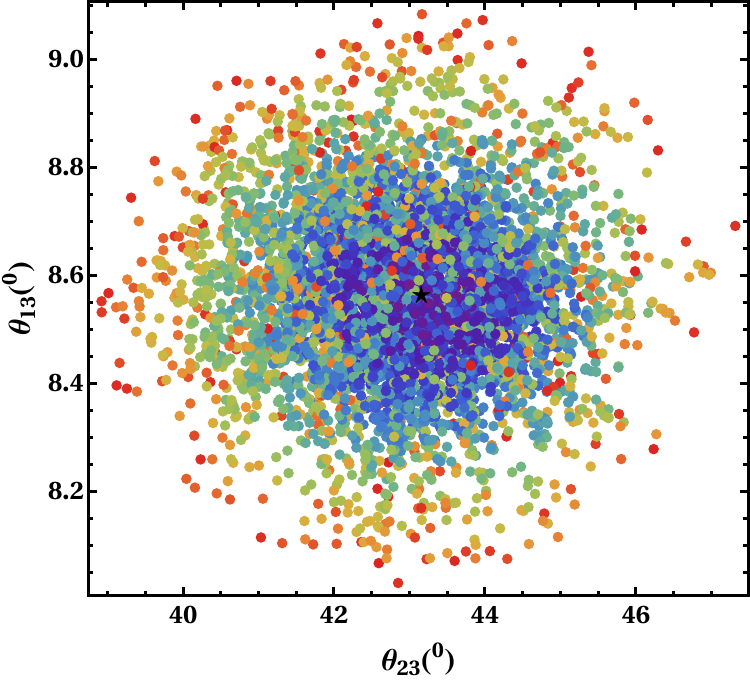}
        \caption{}
        \label{fig3}
    \end{subfigure}
    \hfill
    \begin{subfigure}[b]{0.4\textwidth}
        \centering
        \includegraphics[width=\linewidth]{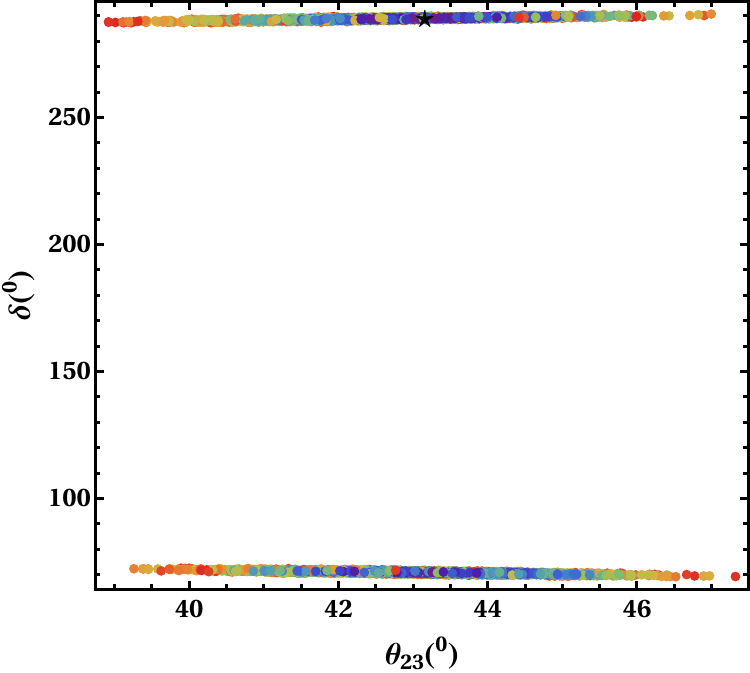}
        \caption{}
        \label{fig4}
    \end{subfigure}
    \hfill
    \begin{subfigure}[b]{0.4\textwidth}
        \centering
        \includegraphics[width=\linewidth]{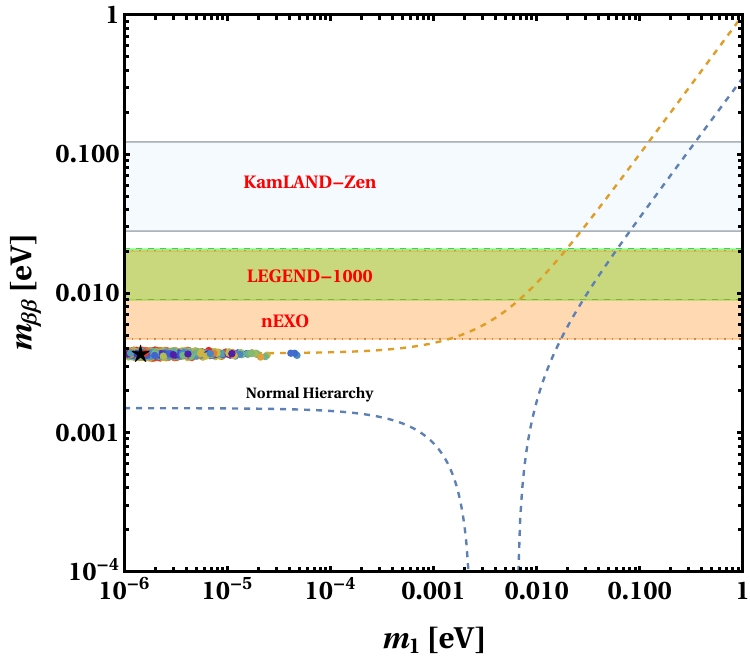}
        \caption{}
        \label{fig5}
    \end{subfigure}
    \hfill
    \begin{subfigure}[b]{0.4\textwidth}
        \centering
        \includegraphics[width=\linewidth]{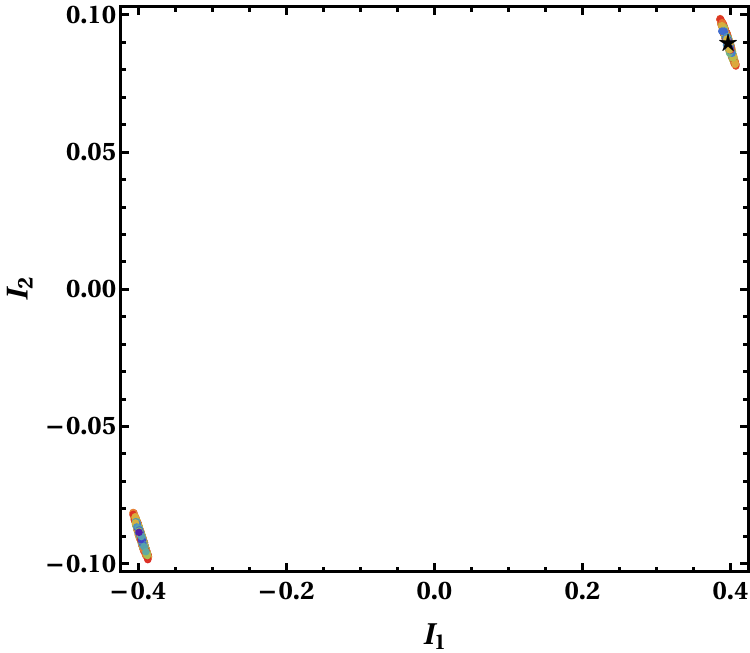}
        \caption{}
        \label{fig6}
    \end{subfigure}
    \vspace{0.2cm}
    \centering
    \includegraphics[width=0.4\linewidth]{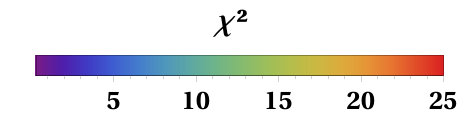}  
    \caption{The allowed parameter space of the complex modulus $\tau$ and predicted ranges of neutrino observables such as mixing angles, Dirac type $CP$ phase, effective Majorana mass, and Majorana CP invariants. The black star ($\star$) corresponds to the best-fit value obtained for $\chi^2_{min}=0.28$.}
    \label{Figure1}
\end{figure}

\begin{figure}[t]
    \centering
 \includegraphics[height=5cm]{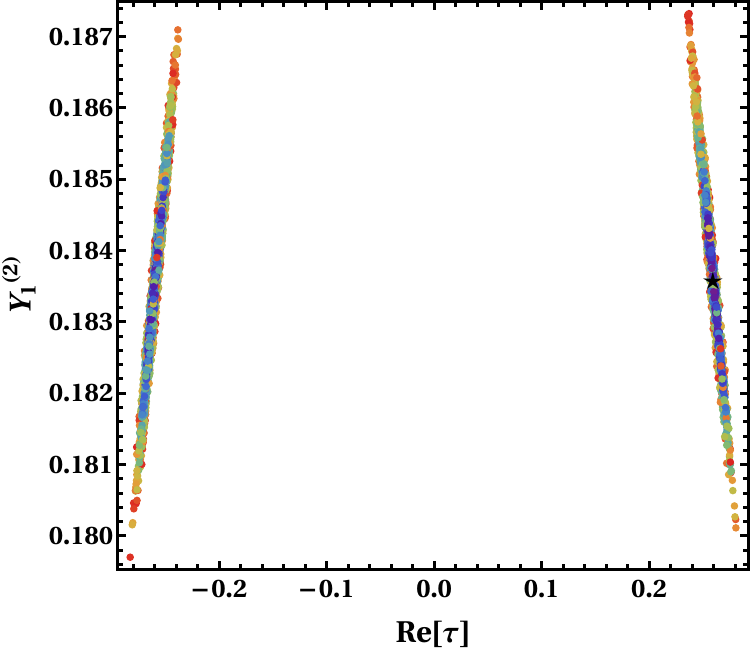}
 \hfill
  \includegraphics[height=5cm]{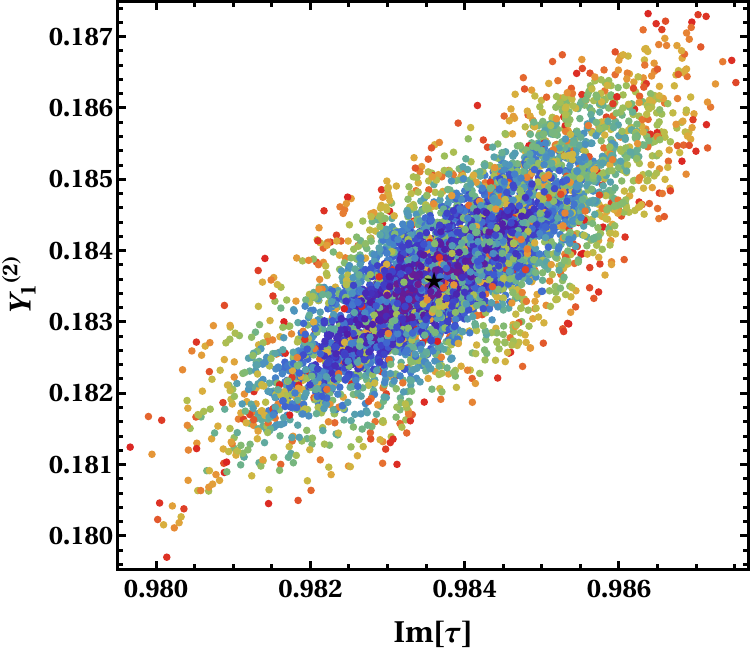}
  \hfill
   \includegraphics[height=5cm]{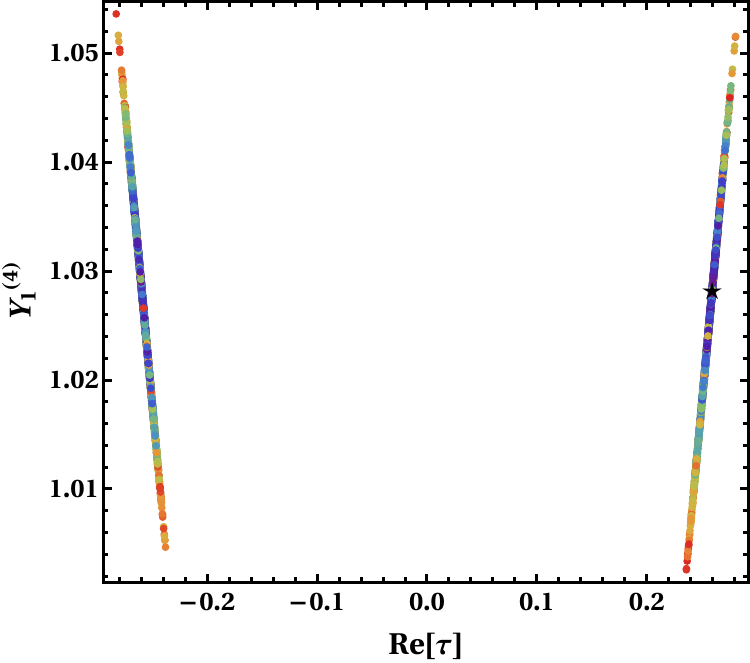}
   \hfill
    \includegraphics[height=5cm]{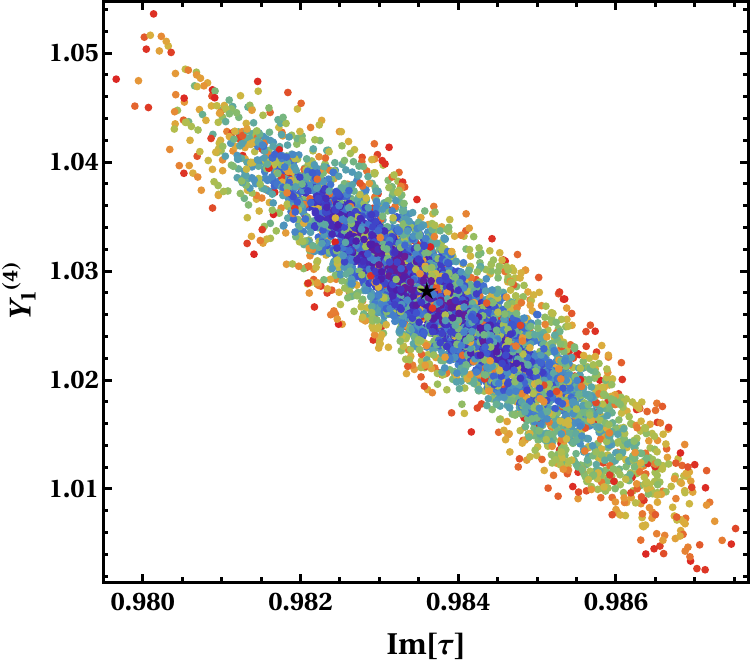}
    \hfill
     \includegraphics[height=5cm]{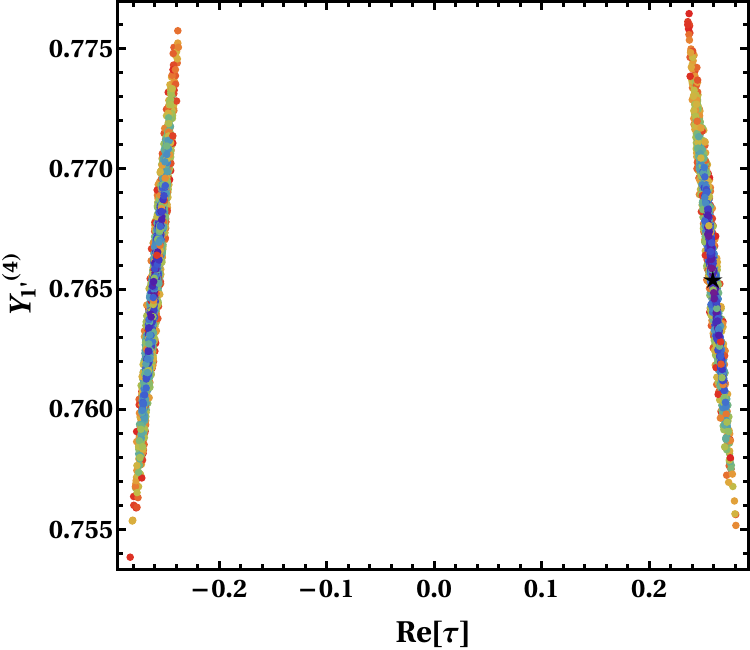}
     \hfill
      \includegraphics[height=5cm]{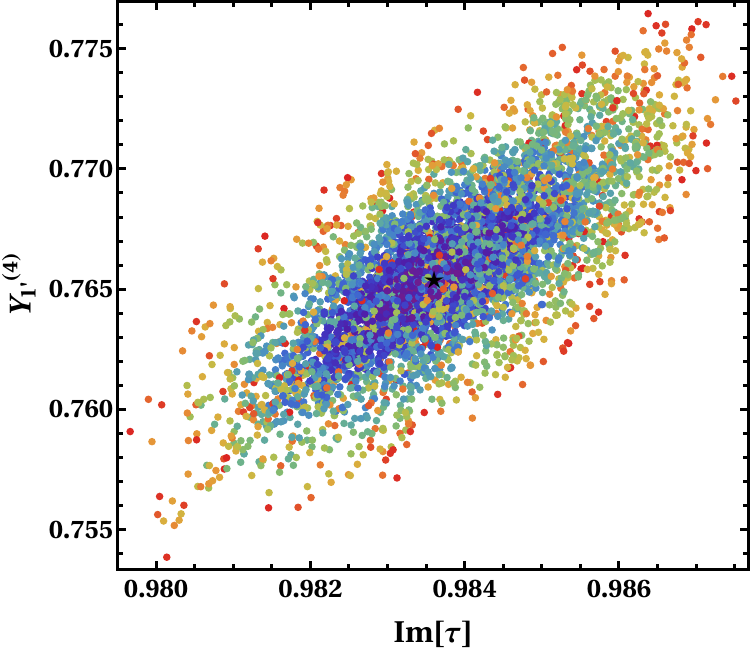}
      \hfill
       \includegraphics[height=5cm]{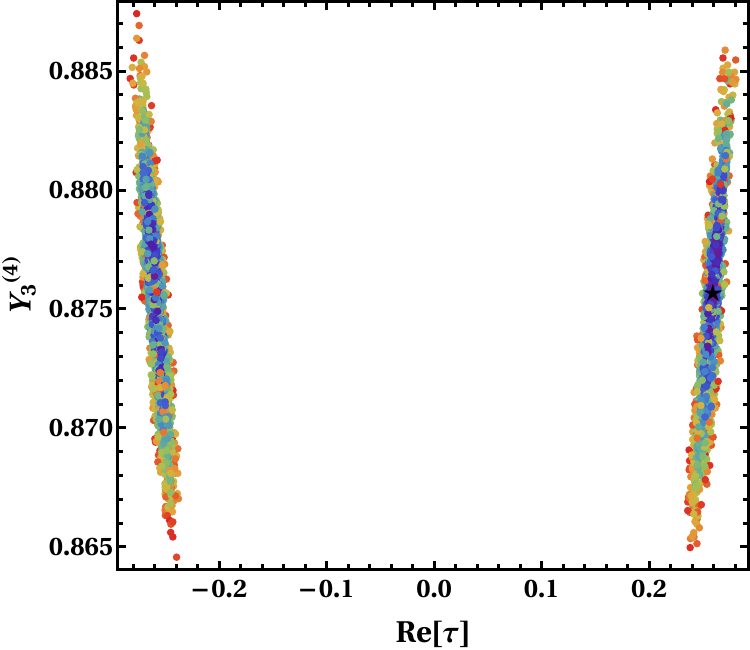}
       \hfill
        \includegraphics[height=5cm]{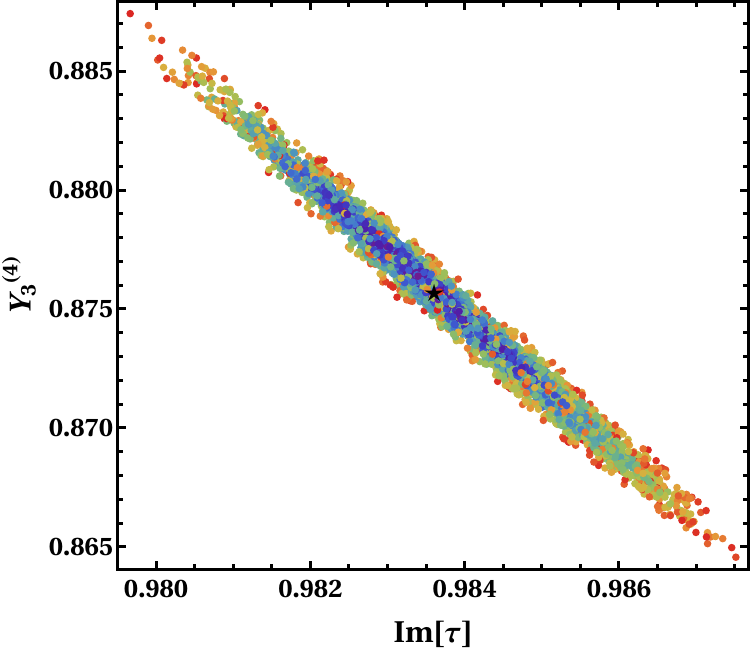}
    \caption{The variation of Yukawa couplings with real and imaginary part of complex modulus $\tau$. The color code is same as for Fig. \ref{Figure1}}
    \label{fig:appendix1}
\end{figure}

\begin{figure}[h!]
 \centering
        \includegraphics[height=6.5cm]{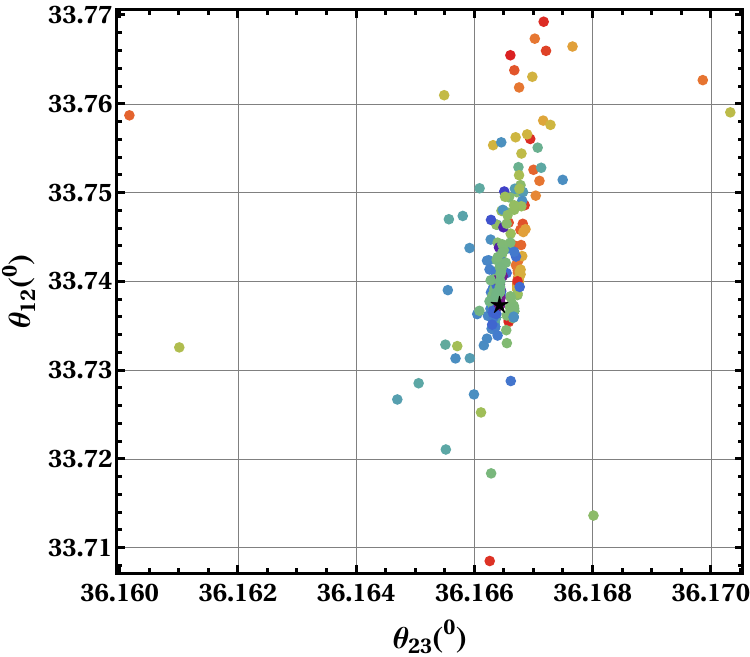}
        \caption{The correlation between atmospheric ($\theta_{23}$) and solar ($\theta_{12}$) mixing angle for IH.}
        \label{figIH}
\end{figure}

\begin{table}[t]
    \centering
    \renewcommand{\arraystretch}{1.6}
    \begin{tabular}{|c|c|}
    \hline
     Input Parameters & Range \\
     \hline \hline
     Re[$\tau$] & $\pm[0.0, 0.5]$ \\ 
     Im[$\tau$] & [0.8, 1.5] \\
     $\gamma$ & $[10^{-6}, 1]$ \\ 
     $\beta_L$ & $[10^{-6}, 1]$ \\ 
     $\kappa_{1,2}$ (GeV) & $[10^{5},10^{10}]$ \\
     $\kappa_s$  (GeV) & $[10^{5},10^{10}]$ \\ 
     $v \alpha_{1}$ (GeV)& $[10^{-3}, 10^{-2}]$\\
 $r_{21}$&$[10^{2}, 1.5 \times 10^{3}]$\\
 $r_{31}$&$[10^{2}, 1.5 \times 10^{3}]$\\
     $m_{\eta_1}$ (GeV) & [500, $10^3$] \\
     $\Delta m^2=m^2_{\eta_I}-m^2_{\eta_1}$ (GeV) & [$10^{-3}$,10]\\
     $\mu_1$ (GeV) & [$10,10^4$]\\
     \hline
    \end{tabular}
    \caption{\centering The ranges of the parameters used for the numerical scan.}
    \label{tab:scan}
\end{table}
The neutrino mass matrix given by Eqn. (\ref{neutrinomassmatrix}) is diagonalized using the relation $U_\nu^T M^\nu U_\nu = \text{diag}(m_{\nu_1}, m_{\nu_2}, m_{\nu_3})$. The mixing matrix $U_{\text{PMNS}}$ = $U_L^\dagger U_\nu$, since the charged lepton mass matrix is not diagonal in the flavor basis. Now, the mixing angle can be extracted from $U_{\text{PMNS}}$ as 
\begin{equation}
    \sin^2 \theta_{13} = |U_{13}|^2, \quad
    \sin^2 \theta_{12} = \frac{|U_{12}|^2}{1 - |U_{13}|^2}, \quad
    \sin^2 \theta_{23} = \frac{|U_{23}|^2}{1 - |U_{13}|^2},
\end{equation}
where $U_{mn}$ are elements of $U_{\text{PMNS}}$. Further, the CP invariants sensitive to Majorana phases can be written in terms of the elements of the PMNS matrix
\begin{equation}
    I_1 = \text{Im}\left[U_{11}^* U_{12}\right],
    \label{I1}
\end{equation}
\begin{equation}
    I_2 = \text{Im}\left[U_{11}^* U_{13}\right].
    \label{I2}
\end{equation}
\noindent Using effective Majorana mass 
\begin{equation}
m_{\beta\beta}
=
\left|
U_{11}^2\, m_1
+
U_{12}^2\, m_2
+
U_{13}^2\, m_3
\right|, 
\end{equation}
the model prediction for $m_{\beta \beta}$ is, also, investigated. We numerically diagonalize the neutrino mass matrix in Eqn. (\ref{neutrinomassmatrix}) and elucidate predictions for the neutrino observables. In order to  ascertain the allowed parameter space, we define the $\chi^2$ function as
\begin{equation}
\chi^2 = \sum_{i = 1}^7 \left( \frac{P_i - P_i^0}{\sigma_i} \right)^2,
     \label{}
\end{equation}
where $P_i$ represents the observables predicted by the model, $P_i^0$ the central value and $\sigma_i$ represents the error corresponding to 1$\sigma$ level as shown in Table \ref{tab:nufit}.
\begin{table}[t]
\small
\centering
\renewcommand{\arraystretch}{1.15}
\begin{tabular}{|l|c|c|c|c|}
\hline
Observable & best-fit$\pm1\sigma$ (NH) & best-fit$\pm1\sigma$ (IH) & $3\sigma$ range (NH) & $3\sigma$ range (IH) \\
\hline \hline
$\theta_{12}~[^{\circ}]$ & $33.8^{+0.7}_{-0.6}$ & $33.8^{+0.7}_{-0.6}$ & $31.3 \to 36.3$ & $31.3 \to 36.3$ \\
$\theta_{23}~[^{\circ}]$ & $49.1^{+0.9}_{-1.0}$ & $49.2^{+0.9}_{-1.0}$ & $39.8 \to 51.9$ & $40.2 \to 52.1$ \\
$\theta_{13}~[^{\circ}]$ & $8.55^{+0.13}_{-0.14}$ & $8.57^{+0.13}_{-0.14}$ & $8.2 \to 8.9$ & $8.2 \to 8.9$ \\
$\Delta m_{21}^2\,[10^{-5}\,\text{eV}^2]$ & $7.42^{+0.21}_{-0.20}$ & $7.42^{+0.21}_{-0.20}$ & $6.82 \to 8.04$ & $6.82 \to 8.04$ \\
$\Delta m_{31}^2\,[10^{-3}\,\text{eV}^2]$ & $+2.510^{+0.027}_{-0.027}$ & $-2.490^{+0.026}_{-0.028}$ & $+2.43 \to +2.58$ & $-2.56 \to -2.42$ \\
$m_e/m_\mu$ & $0.004737$ & $0.004737$ & -- & -- \\
$m_\mu/m_\tau$ & $0.058823$ & $0.058823$ & -- & -- \\
\hline
\end{tabular}
\caption{The neutrino oscillation observables from NuFIT~6.0~\cite{Esteban:2024eli}. The values of the mass ratios of charged leptons have been taken from Ref. \cite{ParticleDataGroup:2024cfk}.}
\label{tab:nufit}
\end{table}
We have sampled the seven observables in this study which include three mixing angles, two mass-squared differences ($\Delta m^2_{21}$ and $\Delta m^2_{31}$), and two lepton mass ratios ($m_e/m_\mu$ and $m_\mu/m_\tau$) (see Table~\ref{tab:nufit}). We explore the parameter space (given in Table \ref{tab:scan}) using $10^{9}$ samples in Monte Carlo simulations with  mass hierarchy constraint $M_{N_1} < M_{f} << M_{N_2}$. The experimental input values used in our analysis are summarized in Table~\ref{tab:nufit}. These parameters demonstrate how well the model reproduces the observed neutrino mixing angles in the case of normal hierarchy (NH) yielding a minimum chi-square value of $\chi^2_{\text{min}} = 0.28$. The model predictions are illustrated through correlation plots obtained for $\chi^2 \leq 25$ where the solid black star denotes the best-fit point corresponding to $\chi^2_{\text{min}} = 0.28$. Fig.~\ref{fig1} presents the allowed region of the complex modulus $\tau = \text{Re}[\tau] + i\; \text{Im}[\tau]$ that remains consistent with the neutrino oscillation data. We find that the data are well fitted with $\text{Re}[\tau] \sim  \pm 0.26$ and $\text{Im}[\tau] \sim  0.97$. The corresponding Yukawa couplings are displayed in Fig.~\ref{fig:appendix1}. The predicted neutrino mixing angles lie within their $3\sigma$ experimental ranges as shown in Fig.~\ref{fig2} and \ref{fig3}. Fig.~\ref{fig4} illustrates the correlation between the atmospheric mixing angle $\theta_{23}$ and the Dirac-type $CP$-violating phase $\delta$. We observe that the model exhibits non-zero CP violation with Dirac phase $\delta$ being highly constrained to lie in the first and fourth quadrants. While NO$\nu$A data allow normal ordering with  $\delta\in[0^\circ,180^\circ]$ and inverted ordering with $\delta\in[180^\circ,360^\circ]$ \cite{NOvA:2021nfi}, the T2K experiment favors values around $\delta\simeq270^\circ$ for both hierarchies \cite{T2K:2023smv}. This region partially overlaps with the NO$\nu$A and T2K-preferred normal ordering solution. The future precision measurements will therefore provide a decisive test for this framework. The predictions for the effective Majorana mass $m_{\beta\beta}$ as a function of the lightest neutrino mass $m_1$ are shown in Fig.~\ref{fig5}. The allowed region corresponds to the parameter space consistent with neutrino oscillation data in the case of NH. The model predicts $m_{\beta\beta}$ values that lie  below the current experimental limits from KamLAND-Zen \cite{KamLAND-Zen:2024eml} and within the future sensitivities of LEGEND-1000 \cite{LEGEND:2025jwu} and nEXO \cite{nEXO:2021ujk}. 
The observation of neutrinoless double beta decay ($0\nu\beta\beta$) in the upcoming experiments with refined sensitivities will therefore provide a critical test of the framework. The correlation between the invariants $I_1$ and $I_2$ (Eqns.~(\ref{I1})–(\ref{I2})) is shown in Fig.~\ref{fig6}. The results indicate small but non-zero values of $I_{1,2}$, implying non-zero Majorana-type $CP$ violation. Furthermore, we have investigated the parameter space for the inverted neutrino mass hierarchy. The model excludes the IH
scenario as illustrated in Fig. \ref{figIH}. The figure illustrates the correlation between the mixing angles $\theta_{12}$ and $\theta_{23}$, showing that $\theta_{23}$ falls outside the 3$\sigma$ range of the global data, thereby, confirming the incompatibility with IH scenario.
While the preference for NH is consistent with global analyses, the prediction of a first-octant solution for $\theta_{23}$ is in mild tension with the slight upper-octant preference reported by T2K and NO$\nu$A. This discrepancy highlights a key testable prediction of our model. The best-fit values of neutrino observables and corresponding model parameters are given in Table \ref{tab:bf_observ} and \ref{tab:bf_para}, respectively. 

\begin{table}[t]
    \centering
    \begin{tabular}{|c| c| c| c| c| c| c| c|}
       \hline
        $\theta_{12} (^\circ)$ & $\theta_{13} (^\circ)$ & $\theta_{23} (^\circ)$ & $\delta (^\circ)$ & $\Delta m^2_{21} (\text{eV}^2)$ & $\Delta m^2_{31} (\text{eV}^2)$ &\(\text{Re}[\tau]\) & \(\text{Im}[\tau]\) \\
        \hline\hline
        34.00 & 8.56 & 43.15 & $289.04$ & $7.5 \times 10^{-5}$ & $2.5 \times 10^{-3}$ & 0.26& 0.98 \\
        \hline
    \end{tabular}
    \caption{The best-fit values for the neutrino oscillation parameters obtained for a minimum $\chi^2$ value of 0.28 for NH.}
    \label{tab:bf_observ}
\end{table}

\begin{table}[t]
    \centering
    \begin{tabular}{|c|c|c|c|}
        \hline
        Parameter & Value & Parameter & Value \\
        \hline \hline
        $\kappa_1$ (GeV) & $7.90\times10^5$ & $\kappa_2$ (GeV) & $6.69\times10^8$ \\
        $\gamma$ & $9.52\times10^{-6}$ & $\kappa_S$ (GeV) & $5.22\times10^8$ \\
        $\beta_L$ & $1.00$ & $m_{\eta_1}$ (GeV) & $6.98 \times 10^2$\\
 $\Delta M_\eta^2$ (GeV)& $9.87 \times 10^{-3}$& $v \alpha_1$ (GeV)&$4.10 \times 10^{-3}$\\
 $r_{21}$& $1.32 \times 10^3$& $r_{31}$&$1.67  \times 10^{2}$\\
 $M_{N_1}$ (GeV)& $1.21 \times 10^6$& $M_{N_2}$ (GeV)&$7.24 \times 10^8$\\
 $M_{f}$ (GeV)& $3.09 \times 10^7$& Br$(\mu \rightarrow e \gamma)$&$5.43 \times 10^{-28}$\\
 \hline
    \end{tabular}
    \caption{The values of the model parameters obtained from the $\chi^2$ analysis corresponding to the $\chi_{min}^2 = 0.28$ for the NH.}
    \label{tab:bf_para}
\end{table}
\section{Lepton Flavor Violation}\label{section4}
One of the key low-energy signatures of physics beyond the SM is the presence of charged lepton flavor-violating (LFV) decays. In Scotogenic frameworks, the presence of new particles involved in the loop can induce sizable contributions to these processes. Among the various LFV channels, the decay $\mu \rightarrow e \gamma$ provides the strongest experimental constraint, with current upper limit $ \mathrm{Br}(\mu \rightarrow e\gamma) < 4.2 \times 10^{-13}$ \cite{Mori:2016vwi}. In our analysis, we impose this bound to restrict the parameter space of the model. The branching ratio for the radiative decay $\mu \to e\gamma$ is \cite{Toma:2013zsa}
\begin{equation}
    \mathrm{BR}(\mu \rightarrow e\gamma)
    = \frac{3(4\pi)^{3}\,\alpha_{em}}{4\, G_{F}^{2}}
    \left| \mathcal{A} \right|^{2}\,
    \mathrm{BR}(\mu \rightarrow e\,\bar{\nu}_{e}\,\nu_{\mu}) ,
    \label{eq:BRmuegamma}
\end{equation}
where $\alpha_{em}$ is the  fine-structure constant,
$G_{F}$
is the Fermi constant, and $\mathrm{BR}(\mu \rightarrow e\,\bar{\nu}_{e}\,\nu_{\mu})$ is approximately equal to unity. The dipole amplitude entering  Eqn. \eqref{eq:BRmuegamma} is
\begin{equation}
    \mathcal{A}
    =\frac{ Y_{11}^*\, Y_{21}}
    {32\pi^{2}\, m_{\eta^{+}}^{2}}
    \; \mathcal{G}(y),
    \label{eq:A1}
\end{equation} 
where $m_{\eta^{+}}$ is the mass of the charged scalar field $\eta^{+}$, 
and $Y_{11}$ and $Y_{21}$ are the corresponding elements of the 
rotated Yukawa coupling matrix in the charged-lepton mass basis. 
This rotated Yukawa matrix is defined as
$
Y = \beta_L U_L^\dagger \,(Y_1^{4},\, Y_3^{4},\, Y_2^{4})^{T},
$
where $U_L$ is the unitary matrix that diagonalizes the charged-lepton mass matrix. Throughout our numerical analysis we fix $m_{\eta^{+}} = m_{\eta_I}\,$.
The loop argument is defined as   $y = \frac{M_{f}^{2}}{m_{\eta^{+}}^{2}}$,
where $M_{f}$ is the mass of the fermion propagating in the loop.

\begin{figure}[t]
 \centering
        \includegraphics[height=6.5cm]{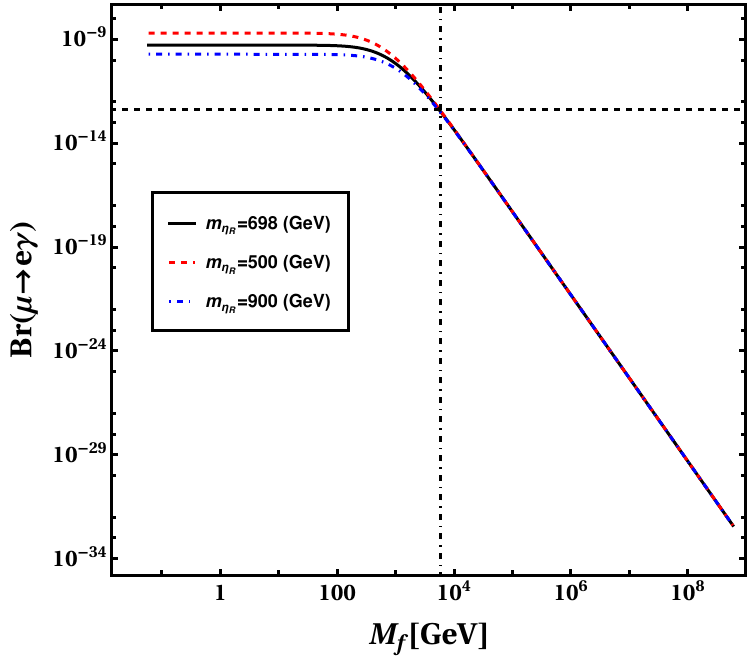}
        \caption{The correlation between mass of dark-sector fermion ($f$) and LFV branching ratio Br$(\mu \rightarrow e \gamma)$ for three different values of scalar mass $m_{\eta_R}\sim m_{\eta_1}$, while keeping all other parameters fixed at their best-fit values given in Table \ref{tab:bf_para}. The horizontal dashed line shows the current experimental upper bound on Br$(\mu \rightarrow e \gamma)$. The vertical dot–dashed line indicates the lower bound on the fermion mass $M_f$ required to satisfy the LFV constraint.}
        \label{fig:Br}
\end{figure}

\noindent The loop function $\mathcal{G}(y)$ appearing in Eqn.~\eqref{eq:A1} is
\begin{equation}
    \mathcal{G}(y)
    =
    \frac{1 - 6y + 3y^{2} + 2y^{3} - 6y^{2} \ln y}
    {6(1-y)^{4}} .
    \label{eq:G1}
\end{equation}
The value of branching ratio predicted at the best-fit point, corresponding to the minimum of the $\chi^{2}$, is presented in Table~\ref{tab:bf_para}. 
To illustrate how $\mathrm{Br}(\mu \rightarrow e\gamma)$ varies with the mass of the fermion $f$ and the real scalar mass $m_{\eta_R}$ circulating in the loop, we fix all other parameters at their best-fit values and compute the resulting correlation for three representative values of $m_{\eta_R}$. The black curve corresponds to the best-fit point, while the other two curves represent the maximum and minimum values of $m_{\eta_R}$ obtained from our data, as shown in Fig.~\ref{fig:Br}.
As seen in the figure, for relatively low fermion masses, up to about $10^{3}\,\mathrm{GeV}$, the suppression is weak and $\mathrm{Br}(\mu \rightarrow e\gamma)$ remains nearly constant and above the current experimental limit. The loop contribution becomes increasingly suppressed, and the branching ratio falls below the experimental upper bound for $M_{f} \gtrsim 6 \times 10^{3}\,\mathrm{GeV}$, irrespective of the mass $m_{\eta_R}$.

\section{Leptogenesis}\label{section5}
In Type-I seesaw framework, successful thermal leptogenesis typically requires a high scale $\mathcal{O}(10^{9})~\text{GeV}$, known as the Davidson--Ibarra bound, to simultaneously account for neutrino masses and the observed baryon asymmetry. To evade this bound and realize leptogenesis at a lower scale, additional CP-violating contributions beyond the minimal seesaw setting are necessary. These contributions are often originates \textit{via} new interactions or loop-induced effects involving extra states in non-holomorohic realization of scoto-seesaw modular framework. To this end, we consider   the participation of dark sector particles in the leptogenesis dynamics, assuming a BSM fermion mass hierarchy   $M_{N_1} < M_{f} <<  M_{N_2}$, for which  the  inverse decays and scatterings involving  $N_2$ are Boltzmann suppressed. Consequently, at temperatures $T \sim M_{N_2}$, the interactions of the heavier fermion $N_2$ remain in thermal equilibrium, leading to an efficient washout of any lepton asymmetry it may generate. Consequently, successful leptogenesis proceeds solely through the out-of-equilibrium decay of $N_1$. For simplicity, we work in the single-flavor approximation, which captures the qualitative behavior of the asymmetry. 
  It is worth noting that the model, also, contains a viable candidate for dark matter. In particular, the CP-even neutral scalar originating from the inert doublet, $\eta_{1}$, can play the role of dark matter provided it is the lightest among  $\kappa$ - odd particle. To ensure that the dark matter is predominantly the doublet component (rather than the singlet $s$), the mixing between the two CP-even states must remain small. This requirement, in Eqn. (\ref{darkmasses}), translates into the conditions
\begin{equation}
A < B, 
\qquad 
|C| \ll |A-B|,
\end{equation}
where
\begin{equation}
A = \mu_{\eta}^2 + \frac{1}{2}(\lambda_3 + \lambda_4 - \lambda_5)\, v^2,
\qquad 
B = \mu_{s}^2 \quad (\text{for } \lambda_8 \to 0),
\qquad 
C = \mu_1\, v.
\end{equation}
Under these conditions, the lighter mass eigenstate is dominantly composed of $\eta_{1}$ and thus serves as the dark matter candidate.
\begin{figure}[t]
\centering
{\begin{tikzpicture}
\begin{feynman}
\vertex (a){\(\rm\bf\color{black}{N_1 }\)};
\vertex [right=2.5cm of a] (b);
\vertex [above right=1.5cm and 1.5cm of b] (c){\(\rm\bf\color{black}{L_{\alpha} }\)};
\vertex [below right=1.5cm and 1.5cm of b] (d){\(\rm\bf\color{black}{H }\)};
\diagram* {
(a) -- [line width=0.25mm,plain,  style=warmblack,ultra thick] (b),
(b)-- [line width=0.25mm,fermion,  style=purple,ultra thick] (c),
(b) -- [line width=0.25mm,charged scalar, style=purple,ultra thick] (d)};
\node at (b)[circle,fill,style=purple,inner sep=1pt]{};
\end{feynman}
\end{tikzpicture}}\quad
{\begin{tikzpicture}
\begin{feynman}
\vertex (a){\(\rm\bf\color{black}{N_1 }\)};
\vertex [right=2.5cm of a] (b);
\vertex [above right=1.5cm and 1.5cm of b] (c);
\vertex [below right=1.5cm and 1.5cm of b] (d);
\vertex [right=1.5cm of c] (e){\(\rm\bf\color{black}{L_{\alpha} }\)};
\vertex [right=1.5cm of d] (f){\(\rm\bf\color{black}{H}\)};
\diagram* {(a) -- [line width=0.25mm,plain,arrow size=1.2pt, style=warmblack,ultra thick] (b),
(b)-- [line width=0.25mm, plain, arrow size=1.2pt, style=black,ultra thick, edge label'={\(\rm\bf\color{black}{f }\)}] (c),
(d) -- [line width=0.25mm,scalar, style=black, ultra thick,arrow size=1.2pt,edge label'={\(\rm\color{black}{s }\)} ] (b),
(c) -- [line width=0.25mm,charged scalar, style=black, ultra thick, arrow size=1.2pt, edge label'={\(\rm\bf\color{black}{\eta }\)}] (d),
(c) -- [line width=0.25mm,fermion,  arrow size=1.2pt, style=purple,ultra thick] (e),
(d) -- [line width=0.25mm,charged scalar, style=purple, ultra thick,arrow size=1.2pt ] (f)};
\node at (b)[circle,fill,style=black,inner sep=1pt]{};
\node at (c)[circle,fill,style=black,inner sep=1pt]{};
\node at (d)[circle,fill,style=black,inner sep=1pt]{};
\end{feynman}
\end{tikzpicture}}
\caption{The tree and one-loop levels Feynman diagrams for dark-sector portal leptogenesis.}
\label{fig:feynman-lepto}
\end{figure}
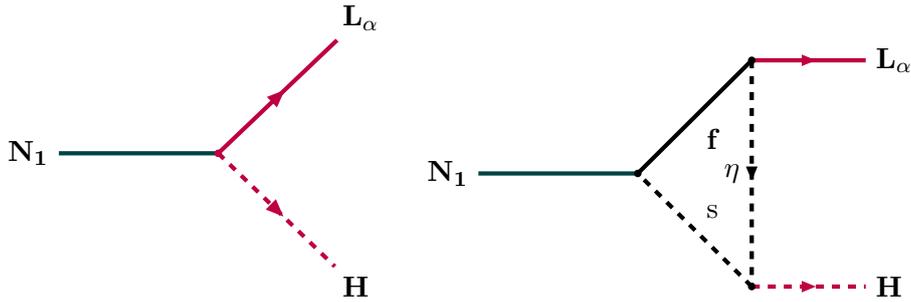
\begin{table}[t]
\scriptsize
    \centering
    \begin{tabular}{|c|c| c| c| c| c| c| c| c|}
       \hline
         Benchmark Point&$M_{N_1}$ (GeV) & $ M_\eta$ (GeV) & $\Delta m^2$ (GeV) & $\mu_s$ (GeV) & $\mu_1$ (GeV) &  $M_{f}$ (GeV) &\(\text{Re}[\tau]\) & \(\text{Im}[\tau]\) \\
        \hline\hline
        BP1&$1.21 \times 10^6$ & $6.98 \times 10^2$  &$9.87 \times 10^{-3}$  & $10^3$ & $10^4$ &$3.09 \times 10^7$  &$0.26$ &$0.98$  \\
        BP2&$1.28 \times 10^6$ & $6.02 \times 10^2$&$9.97 \times 10^{-3}$   & $10^3$ &  $5\times10^3$&$3.40 \times 10^7$  &$0.25$ &  $0.98$\\
            BP3&$1.23 \times 10^6$ &$6.23 \times 10^2$  &$9.96 \times 10^{-3}$  &$10^3$  & $10^3$ & $3.24 \times 10^7$ &$0.25$ & $0.98$ \\
        \hline
    \end{tabular}
    \caption{The best-fit values for the neutrino oscillation parameters obtained from the $\chi^2$ analysis correspond to a minimum $\chi^2$ value of 0.28 for NH.}
    \label{tab:bflep}
\end{table}

\noindent The CP asymmetry,
\begin{equation}
\epsilon_1 = \frac{\Gamma (N_1 \rightarrow L H)-\Gamma (N_1 \rightarrow \overline{L} H^\dagger)}{\Gamma (N_1 \rightarrow L H)+\Gamma (N_1 \rightarrow \overline{L} H^\dagger)} = \frac{\Gamma (N_1 \rightarrow L H)-\Gamma (N_1 \rightarrow \overline{L} H^\dagger)}{\Gamma_N},
\end{equation}
generated in the decay of $N_1$, shown in Fig.~\ref{fig:feynman-lepto}, arises from the interference between the tree-level and one-loop diagrams 
which, further, can be written as \cite{Barreiros:2022fpi,Borah:2025hpo}
\begin{equation}
\epsilon_1 = \frac{1}{8\pi} \frac{\text{Im}[\mathcal{M}_{\text{tree}}^* \mathcal{M}_{\text{loop}}]}{|\mathcal{M}_{\text{tree}}|^2},
\end{equation}
\begin{align}
\epsilon_1 &= \frac{1}{8 \pi}\frac{{\rm Im}(y_D^\dagger y_f y_1 \mu_1)}{(y_D^\dagger y_D)M_{N_1} (1-\vartheta+\omega)\sqrt{(1-\vartheta+\omega)^2-4\omega}} I_{\text{loop}},
\label{eq:cp_asymmetry_full}
\end{align}
where 
\begin{eqnarray}
I_{\text{loop}}= 1+\vartheta -2\sqrt{\varsigma} \nonumber +(\varkappa-\sqrt{\varkappa}(1-\vartheta+\omega)-\varrho+\omega) {\rm ln}\left[ \frac{\varkappa-(1-\sqrt{\varsigma})^2}{\varkappa-\varsigma+\vartheta}\right],
\end{eqnarray}
is the loop function arises from vertex corrections involving the dark-sector fermion $f$, inert scalars and $y_D=U^{\dagger}_L M_D/v$. The mass ratios are defined as,
\begin{equation}
\varkappa =\frac{M_f^2}{M_{N_1}^2}, \quad \varrho=\frac{m_\eta^2}{M_{N_1}^2}, \quad \vartheta=\frac{m_h^2}{M_{N_1}^2}, \quad \omega=\frac{m_l^2}{M_{N_1}^2}, \quad \varsigma=\frac{m_s^2}{M_{N_1}^2}.
\end{equation}
All CP-violating phases entering $\epsilon_1$ originate from the complex modular forms. With the modular symmetry constraints, the Yukawa couplings in our model are determined by modular forms
\begin{equation}
y_f =\beta_L Y_3^{(4)}, \quad y_1 = a Y_3^{(2)}.
\end{equation}
In the limit of vanishing SM Higgs and lepton masses, the CP asymmetry simplifies to
\begin{align}
\epsilon_1=\frac{1}{8 \pi}\frac{{\rm Im}(y_D^\dagger y_f y_1 \mu_1)}{(y_D^\dagger y_D)M_{N_1}} \left(1-2\sqrt{\varsigma}+(\varkappa-\sqrt{\varkappa}-\varrho){\rm ln}\left[ \frac{\varkappa-(1-\sqrt{\varsigma})^2}{\varkappa-\varsigma}\right] \right).
\label{eq:cp_asymmetry_simple}
\end{align}
The corresponding Boltzmann equations for co-moving densities of $N_1$ and $B-L$ can be written as
\begin{align}
    \frac{dY_{\rm N_1}}{dz} =  & -D_{\rm N}\left(Y_{\rm N_1}-Y_{\rm N_1}^{\rm eq}\right) - \frac{s}{\rm H(z)z} \left(Y_{\rm N_1}^{2}-\left( Y_{\rm N_1}^{\rm eq} \right)^{2}\right)  \bigg [\langle \sigma v \rangle_{N_1 N_1 \longrightarrow s s} + \langle \sigma v \rangle_{N_1 N_1 \longrightarrow H H^{\dagger}} \nonumber \\
    & + \langle \sigma v \rangle_{N_1 N_1 \longrightarrow L_{\alpha} \overline{L}_{\beta}}\bigg ]- \frac{s}{\rm H(z)z} \left(Y_{\rm N_1}-Y_{\rm N_1}^{\rm eq}\right) \bigg [2Y_{l}^{\rm eq}\langle \sigma v \rangle_{\overline{l} N_1 \longrightarrow \overline{q}t} + 4Y_{t}^{\rm eq}\langle \sigma v \rangle_{ N_1 t \longrightarrow \overline{l}q} \nonumber 
      \\
     & + 2Y_{s}^{\rm eq}\langle \sigma v \rangle_{N_1 s \longrightarrow \overline{L} \eta^{\dagger}} + 2Y_{H}^{\rm eq}\langle \sigma v \rangle_{N_1 H \longrightarrow l_{\alpha} V_{\mu}}+ 2y_{f}^{\rm eq}\langle \sigma v \rangle_{N_1 f \longrightarrow H \eta^{\dagger}}\bigg ],
\end{align}
\begin{align}
    \frac{dY_{\rm B-L}^{}}{dz} = & - \epsilon_{1}D_{\rm N}\left(Y_{\rm N_1}-Y_{\rm N_1}^{\rm eq}\right) - W_{\rm ID} Y_{\rm B-L}
     - \frac{\rm s}{\rm H(z)z} Y_{\rm B-L} \Big[ 2Y_{\rm H}^{\rm eq} \langle \sigma v \rangle_{l H^{\dagger}\longrightarrow \overline{L} H} 
     +Y_{\rm N_1} \langle \sigma v \rangle_{\overline{L}N_1 \longrightarrow \overline{q}t} \nonumber \\
     & + 2 Y_q^{\rm eq} \langle \sigma v \rangle_{\overline{L}q \longrightarrow N_1 t} + 2 Y_{l}^{\rm eq} \langle \sigma v  \rangle_{L L \longrightarrow H^{\dagger}H^{\dagger}} + Y_{\eta}^{\rm eq} \langle \sigma v  \rangle_{\overline{L} \eta^{\dagger} \longrightarrow N_1 s } + Y_{V}^{\rm eq} \langle \sigma v \rangle_{\overline{L} V_{\mu}\longrightarrow H N_1} \Big].
\end{align}
Here $Y_i=n_i/s$ denotes comoving density with $n_i$ being number density of species `$i$' and $\rho_s=\frac{2\pi^2}{45}g_{*} T^3$ being entropy density of the Universe. $z=M_{N_1}/T$ and $H(z) =\sqrt{\frac{4\pi^3 g_*}{45}}\frac{T^2}{M_{\rm Pl}}$ is the Hubble parameter at high temperatures. The decay term $D_N$ is defined as
\begin{eqnarray}
D_N  =  \dfrac{ \langle \Gamma_{N} \rangle}{{H} z} = K_{N}z\dfrac{\kappa_{1}(z)}{\kappa_{2}(z)},
\end{eqnarray}
where $K_N=\Gamma_N/{ H}(z=1)$ with $\kappa_i(z)$ being the modified Bessel function of $i^{th}$ kind. The washout due to inverse decay (ID) is
\begin{eqnarray}
W_{\rm ID} =  \dfrac{1}{4}K_{N}z^{3}\kappa_{1}(z).
\end{eqnarray}
The thermal averaged cross-section is defined as
\begin{equation}
    \langle \sigma v \rangle_{ij \rightarrow kl} = \frac{1}{8Tm^2_i m^2_j \kappa_2 (z_i) \kappa_2 (z_j)} \int^{\infty}_{(m_i+m_j)^2} ds \frac{\lambda (\rho_s, m^2_i, m^2_j)}{\sqrt{\rho_s}} \kappa_1 (\sqrt{\rho_s}/T) \sigma,
\end{equation}
with $z_i=m_i/T$ and $\lambda (\rho_s, m^2_i, m^2_j)=[\rho_s-(m_i+m_j)^2][\rho_s-(m_i-m_j)^2]$. The lepton asymmetry at the sphaleron decoupling epoch $T_{\rm sph} \sim 130$ GeV gets converted into baryon asymmetry as
\begin{align}
& Y_B\simeq a_\text{Sph}\,Y_{B-L}=\frac{8\,N_F+4\,N_H}{22\,N_F+13\,N_H}\,Y_{B-L} =0.34 Y_{B-L}\,, 
\label{eq:sphaleron}
\end{align}
with $N_F=3\,,N_H=2$ being the fermion generations and the number of scalar doublets in our model, respectively.

\begin{figure}[t]
 \centering
        \includegraphics[height=6.5cm,width=7cm]{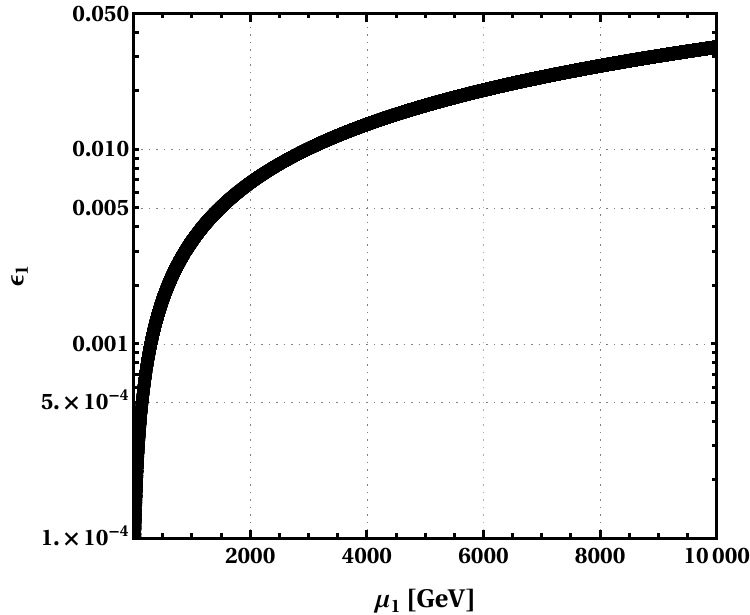}
        \caption{The correlation between CP asymmetry ($\epsilon_1$) as function of dimensionful coupling $\mu_1$ keeping other parameters fixed using Table \ref{tab:bf_observ}.}
        \label{fig:CPasy}
\end{figure}

\begin{figure}[t]
 \centering
         \includegraphics[height=6.5cm,width=7cm]{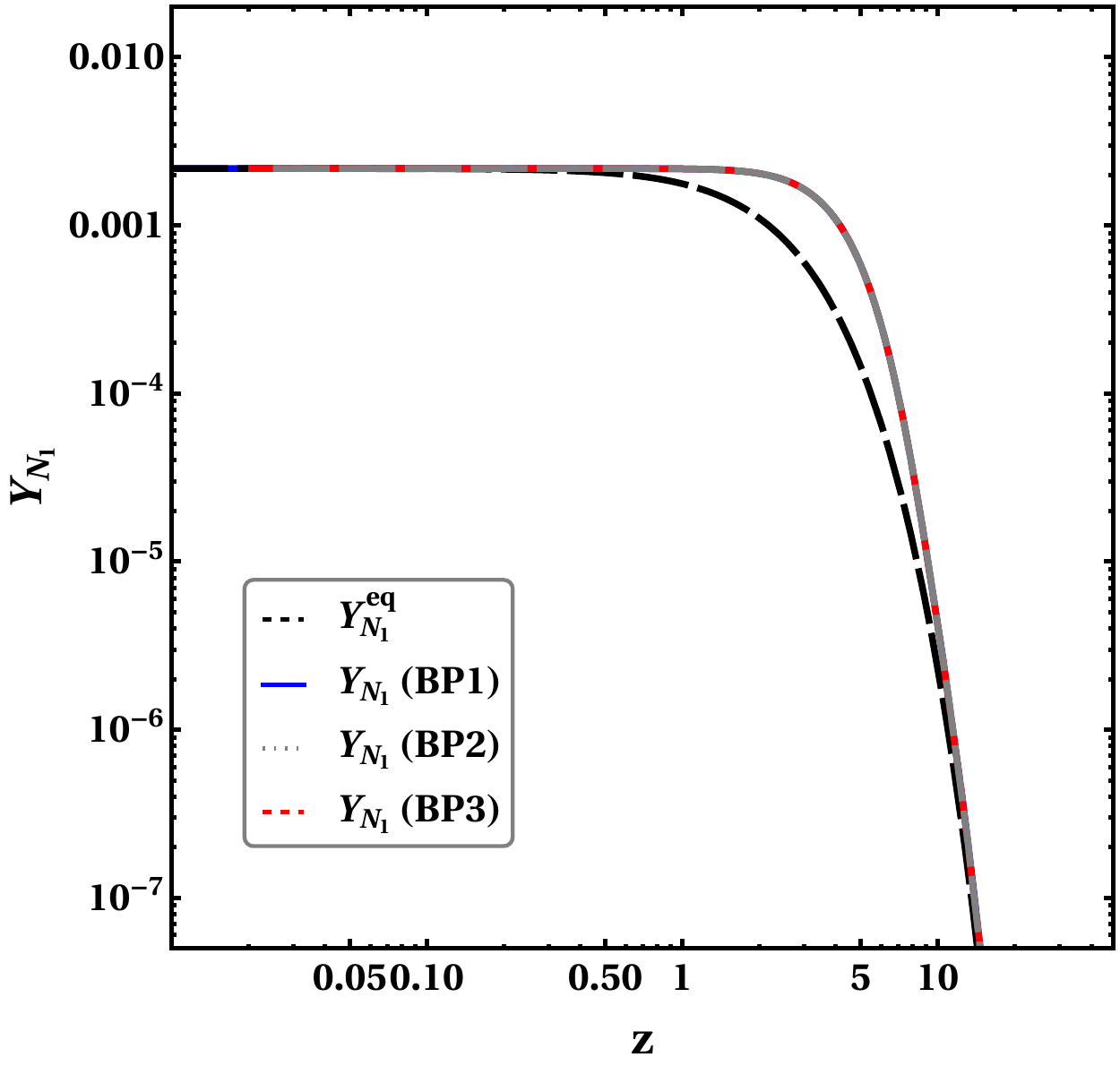}
        \includegraphics[height=6.5cm,width=7cm]{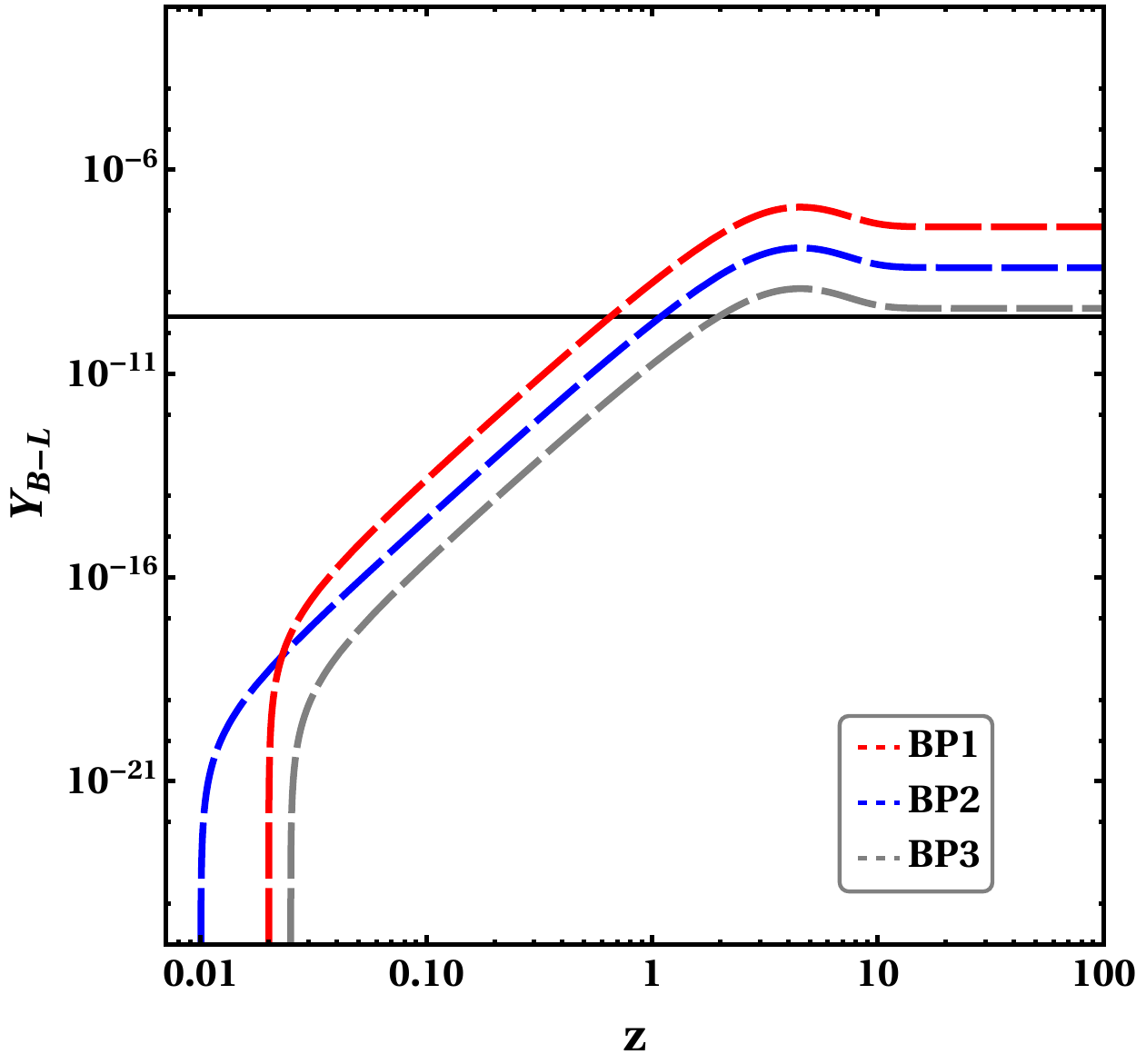}
        \caption{\textbf{Left panel}: The evolution of comoving number density of $N_1$ for different benchmark points given in Table \ref{tab:bflep}. \textbf{Right panel}: The evolution of comoving number density of $B-L$ asymmetry for different benchmark points. The solid horizontal line corresponds to the experimentally required value of the lepton asymmetry (using Eqn. \ref{eq:sphaleron}).}
        \label{fig:La}
\end{figure}
 \noindent The dependence of the CP asymmetry parameter $\epsilon_1$ on the trilinear coupling $\mu_{1}$ is shown in Fig.~\ref{fig:CPasy} where all other parameters are fixed according to Table~\ref{tab:bf_observ}. We observe that increasing $\mu_{1}$ leads to a substantial enhancement of the asymmetry parameter $\epsilon_{1}$ as $\epsilon_1\propto\mu_1$ Eqn.~(\ref{eq:cp_asymmetry_simple}). Though increasing $\mu_{1}$ enhances the CP asymmetry, for sufficiently large values of $\mu_{1}$ (e.g.\ $\mu_{1} \sim 10^{4}\,\text{GeV}$) the resulting $\epsilon_{1}$ can become as large as $\mathcal{O}(1)$, which in turn leads to a $B\!-\!L$ asymmetry exceeding the value required to reproduce the observed baryon asymmetry. This indicates that $\mu_{1}$ cannot be arbitrarily increased, and successful leptogenesis imposes an upper bound on the trilinear coupling $\mu_{1}$ to prevent the overproduction of the baryon asymmetry. The numerical results for the comoving number density of $N_1$ (left panel) and the $B-L$ asymmetry (right panel) are presented in Fig. \ref{fig:La}. With the evolution of the Universe or increase in $z$, the $N_1$ starts decaying to the SM particles (Fig. \ref{fig:feynman-lepto}), therefore reducing their respective comoving number densities. Fig. \ref{fig:La} (left panel) shows the evolution of the comoving number density of $N_{1}$ as a function of 
$z = M_{N_{1}}/T$ for the three benchmark points considered. At early times ($z \ll 1$), all solutions closely track the equilibrium distribution $Y_{N_{1}}^{\text{eq}}$, as expected for a relativistic species efficiently coupled to the thermal bath. As the temperature drops below $M_{N_{1}}$, the equilibrium abundance decreases, while the actual abundance departs from equilibrium and begins to freeze out.

\noindent For all benchmark points, the number density $Y_{N_{1}}$ remains slightly above the equilibrium value during the freeze-out phase. This mild departure from equilibrium reflects the fact that the decay and scattering interactions of $N_{1}$ are still moderately efficient around $z \sim \mathcal{O}(1\!-\!10)$, delaying the freeze-out relative to the purely equilibrium expectation. The three benchmark curves overlap throughout the evolution, indicating that the corresponding variations in the Yukawa couplings or dark-sector parameters have only a sub-leading impact on the dynamics of $N_{1}$ depletion. As a consequence, the CP-asymmetry–producing decays take place in a regime where $N_{1}$ is slightly over-abundant compared to equilibrium, ensuring a non-zero departure from equilibrium. 

\noindent The evolution of $B\!-\!L$ asymmetry for three benchmark points (Table~\ref{tab:bflep}) is shown in right panel of  Fig.~\ref{fig:La}. In all cases, the asymmetry begins to build up once the decays of $N_1$ start to dominate over inverse decay and scattering washout processes. Since the Dirac Yukawa coupling is fixed to a small value $y_D \sim 10^{-6}$ by neutrino oscillation data, the inverse-decay washout term, which scales as $\Gamma_{\rm ID} \propto y_D^2$, remains weak throughout the relevant temperature range. Consequently, the washout rate $\Gamma_{\rm ID}/H$ becomes inefficient at relatively early times, allowing the generated CP asymmetry to accumulate smoothly.

\noindent A distinctive feature of this framework is that the dark-sector responsible for CP violation is controlled by the trilinear parameter $\mu_1$ while the effective washout involving dark-sector fields depends on the fixed parameter $\mu_s = 1~\mathrm{TeV}$. For simplicity, $\mu_s$ is kept the same for all benchmark points, such that the $s$-mediated washout channels remain moderately suppressed. Moreover, the kinematic hierarchy $M_{N_1} \sim 10^6~\mathrm{GeV} < M_f \sim 10^7~\mathrm{GeV}$ forbids on-shell $N_1N_1 \rightarrow ff$ scatterings thereby eliminating an otherwise potentially strong source of thermal washout. The remaining scattering channels, such as $N_1N_1 \rightarrow HH^\dagger$ and $N_1N_1 \rightarrow L\overline{L}$, are further suppressed due to their dependence on the small Yukawa coupling $y_D$ and the relatively small inert-scalar mass splitting $\Delta m^2$.

\noindent The benchmark points in Table~\ref{tab:bflep} correspond to progressively smaller values of $\mu_1$, leading to different magnitudes of the loop-induced CP asymmetry. As a result, BP1 produces the largest final asymmetry, followed by BP2 and BP3, as reflected in Fig.~\ref{fig:La} (right panel). In all benchmarks, the $B\!-\!L$ asymmetry saturates around $z \sim 10$ corresponding to the epoch at which $N_1$ fully departs from equilibrium and washout effects become negligible. The resulting asymmetries are compatible with the value required to reproduce the observed baryon asymmetry of the Universe after sphaleron conversion, demonstrating that the model can successfully realize leptogenesis without invoking resonant enhancement.
 The lightest dark sector particle, $ i.e.$ real part of inert scalar, is our viable dark matter particle (WIMP type) in the model which has been widely explored in the literature \cite{Mandal:2021yph,Borah:2025hpo}. A distinctive prediction of our construction arises from the inert scalar sector, in which the charged scalar $\eta^\pm$ is nearly degenerate with the dark matter candidate $\eta_1$. For the benchmark point, with $\delta m=m_{\eta^\pm} - m_{\eta_1} \sim 0.1$~GeV, the decay 
$\eta^\pm \to \eta_1 + W^{\pm *}$ proceeds only through an off-shell $W$ boson and is therefore phase-space suppressed. Since $m_{\eta^\pm} - m_{\eta_R} \ll m_W$, the resulting decay width scales as $\Gamma \propto (\delta m)^5$ leading to a macroscopic decay length of 
$c\tau_{\eta^\pm} \sim \mathcal{O}(1\!-\!10)\,$cm. 
Consequently, $\eta^\pm$ behaves as a long-lived charged particle inside ATLAS/CMS experiments. The ATLAS Collaboration has recently performed a dedicated search for long-lived charged states using the full $140~\text{fb}^{-1}$ dataset of $pp$ collisions at $\sqrt{s}=13~\text{TeV}$. The analysis exploits signatures characterized by high specific ionization energy loss and time-of-flight measurements \cite{ATLAS:2025fdm}. Interpreting their results\footnote{The same interpretation has been inferred in Ref. \cite{Guo:2025xmz}.} in the context of stau-to-gravitino decays, ATLAS excludes masses up to 560 GeV for such long-lived charged particles. Observation of such a long-lived charged particle signal would provide strong support for the inert scalar sector of the model and offer an important experimental handle on the dark-sector–assisted origin of neutrino masses and leptogenesis.

\section{Conclusion}\label{section6}
This work investigates the neutrino phenomenology and dark-sector assisted leptogenesis in a scoto-seesaw framework constructed under a non-holomorphic $A_4$ modular flavor symmetry. To ensure a clean separation between the fields responsible for the tree-level (seesaw) and loop-induced (scotogenic) contributions to the neutrino mass matrix, we assign even modular weights to the fields in the seesaw sector and odd modular weights to those in the radiative sector. This modular-weight arrangement forbids undesired mixing terms and enforces a controlled hierarchical structure between these two neutrino mass–generating sectors. By restricting the complex modulus $\tau$ to lie within its fundamental domain and performing a systematic numerical scan over the remaining model parameters, we identify regions of parameter space that are fully consistent with the latest neutrino global-fit data. We find that the data is satisfied for $\text{Re}[\tau] \sim  \pm 0.26$ and $\text{Im}[\tau] \sim  0.97$. Also, the CP violating phase $\delta$ is lying $\sim70^{\circ}$ and $\sim290^{\circ}$. The model is consistent only with a normal hierarchical neutrino mass pattern. In this setup, the complex modulus $\tau$ acts as the unique source of CP violation, since all Yukawa couplings and scalar-sector parameters are taken to be real. Consequently, the predicted low-energy neutrino mixing phases and the high-energy CP-violating interactions, relevant for leptogenesis, are entirely determined by the modular origin of flavor sector. Within this construction, successful leptogenesis emerges from the CP-violating out-of-equilibrium decays of the lightest right-handed neutrino $N_1$ into SM leptons and the Higgs doublet at $M_{N_1}\sim10^6$ GeV. The loop-induced CP asymmetry is generated through one-loop diagrams mediated by the dark-sector fields intrinsic to the scotogenic mechanism, notably, without relying on any mass degeneracy or resonance enhancement between the heavy neutrino states. This demonstrates that the interplay of modular flavor structures and radiative dynamics can, naturally, accommodate the observed baryon asymmetry of the Universe while simultaneously reproducing the correct neutrino mass and mixing pattern. Finally, we have, also, incorporated a comment on the unique collider signature of $\eta^\pm$ which may behave as a long-lived charged particle at ATLAS/CMS experiments.

\section*{Acknowledgments}
\vspace{.3cm}
\noindent The work of SN is supported by the United Arab Emirates University (UAEU) under UPAR Grant No. 12S162. LS acknowledges the financial support provided by the Council of Scientific and Industrial Research (CSIR) vide letter No. 09/1196(18553)/2024-EMR-I. Tapender acknowledges the financial support provided by Central University of Himachal Pradesh in the form of freeship. The authors, also, acknowledge Department of Physics and Astronomical Science for providing necessary facility to carry out this work.

\appendix
\section{Non-Holomorphic Modular Flavor Symmetry}
\label{sec:non-holomorphic-modular}
The conventional modular flavor symmetry employs holomorphic modular forms as Yukawa couplings, a structure naturally emerging in supersymmetric settings. However, modular invariance alone does not necessitate holomorphicity. By relaxing this condition, we enter the broader framework of \emph{non-holomorphic modular symmetry}, where Yukawa couplings are polyharmonic Maass forms - automorphic functions satisfying the harmonic condition
\begin{equation}
\Delta_k Y(\tau)=0,
\end{equation}
with $\Delta_k=-4y^2\partial_\tau\partial_{\bar\tau}+2iky\partial_{\bar\tau}$ being the weight-$k$ hyperbolic Laplacian, rather than requiring holomorphicity. These functions transform covariantly under the modular group:
\begin{equation}
Y(\gamma\tau)=(c\tau+d)^k\rho(\gamma)Y(\tau),\qquad\gamma\in\mathrm{SL}(2,\mathbb{Z}),
\end{equation}
where $\rho$ denotes a representation of the finite modular group $\Gamma'_N=\mathrm{SL}(2,\mathbb{Z})/\Gamma(N)$ (or $\Gamma_N$ for even weight). Their Fourier expansion reveals the characteristic non-holomorphic structure:
\begin{equation}
Y(\tau)=\sum_{n\ge0}c^+(n)q^n+c^-(0)y^{1-k}+\sum_{n<0}c^-(n)\Gamma(1-k,-4\pi n y)q^n,
\end{equation}
where the terms proportional to $c^-(0)$ and $c^-(n)$ explicitly break holomorphicity.

\noindent A crucial connection to ordinary modular forms exists via differential operators: the Maass raising operator $D^{1-k}$ and the $\xi$-operator $\xi_k=2iy^k\overline{\partial_{\bar\tau}}$ map weight-$k$ polyharmonic Maass forms to weight-$(2-k)$ modular forms in conjugate and original representations, respectively. This enables explicit construction of polyharmonic forms from known modular forms through the relations
\begin{align}
\xi_k Y^{(k)}_{\mathbf r}(\tau) &= \alpha\, \Omega\, Y^{(2-k)}_{\mathbf r'}(\tau), \\
D^{1-k} Y^{(k)}_{\mathbf r}(\tau) &= \beta\, Y^{(2-k)}_{\mathbf r}(\tau),
\end{align}
where $\Omega$ relates conjugate representations and $\alpha,\beta$ are normalization constants.

\noindent In flavor model building, matter fields $\psi_i$ are assigned modular weights $-k_i$ and transform in irreducible representations $\rho_i$ of $\Gamma'_N$. Yukawa couplings are now polyharmonic Maass forms $Y^{(k_Y)}_{\mathbf r}(\tau)$ of weight $k_Y$ and representation $\mathbf r$. Modular invariance imposes the conditions
\begin{align}
k_Y &= k_{\psi^c}+k_\psi+k_H, \\
\rho_Y\otimes\rho_{\psi^c}\otimes\rho_\psi\otimes\rho_H &\ni \mathbf{1}.
\end{align}

\noindent This non-holomorphic extension significantly expands the Yukawa coupling space while preserving modular invariance. Negative modular weights become admissible, and non-holomorphic terms introduce new parameters that enhance phenomenological flexibility. The vacuum expectation value of $\tau$ simultaneously breaks modular and CP symmetries, offering a unified origin for flavor structures and CP violation. As an illustrative example, at level $N=3$ (finite group $A_4$), polyharmonic Maass forms exist for weights $k=-4,-2,0,2,4,6,\ldots$, organizing into $A_4$ multiplets, thereby enabling novel texture constructions beyond the holomorphic paradigm.

\bibliographystyle{JHEP}
\bibliography{ref}

\end{document}